\documentclass[12pt]{article}
\usepackage{amsmath}
\begin{document}
\def\thefootnote{\fnsymbol{footnote}}
\begin{flushright}
KANAZAWA-10-11  \\
November, 2010
\end{flushright}
\vspace{ .7cm}
\vspace*{2cm}
\begin{center}
{\LARGE\bf Affleck-Dine leptogenesis in the radiative neutrino mass model}\\
\vspace{1 cm}
{\Large H.~Higashi}\footnote{e-mail:~higa@hep.s.kanazawa-u.ac.jp},
{\Large T.~Ishima}\footnote{e-mail:~ishima@hep.s.kanazawa-u.ac.jp}
{\Large and D.~Suematsu}\footnote{e-mail:~suematsu@hep.s.kanazawa-u.ac.jp},
\vspace*{1cm}\\
{\itshape Institute for Theoretical Physics, Kanazawa University,\\
        Kanazawa 920-1192, Japan}\\
\end{center}
\vspace*{1cm}
{\Large\bf Abstract}\\
Radiative neutrino mass models have interesting features,
which make it possible to relate neutrino masses to the existence 
of dark matter.
However, the explanation of the baryon number asymmetry in the universe 
seems to be generally difficult as long as we suppose leptogenesis 
based on the decay of thermal right-handed neutrinos.
Since right-handed neutrinos are assumed to have masses of 
$O(1)$ TeV in these models, they are too 
small to generate the sufficient lepton number asymmetry.  
Here we consider Affleck-Dine leptogenesis in a radiative neutrino mass
model by using a famous flat direction 
$LH_u$ as an alternative possibility. The constraint on the reheating
temperature could be weaker than the ordinary models.
The model explains all the origin of the neutrino masses, 
the dark matter, and also the baryon number asymmetry in the universe. 
\newpage
\section{Introduction}
Recent observations of the existence of neutrino masses \cite{nmass} and 
dark matter (DM) \cite{wmap} are crucial 
ingredients to consider physics beyond the standard model (SM).
The origin of the baryon number asymmetry in the universe also remains
an unsolved problem in the SM \cite{basym}. 
Both the neutrino masses and the baryon number asymmetry are known to be 
explained in a unified way through the leptogenesis scenario 
in the framework of the seesaw mechanism \cite{leptg1}. 
Extensive studies on this subject have been done during recent 
 years \cite{leptg2}.
On the other hand, supersymmetry can play a crucial role for the 
explanation of the DM abundance in the universe \cite{mssm}, although it
has been introduced originally to solve the hierarchy problem.
Supersymmetric models have a good candidate for the DM as
the lightest superparticle (LSP) as long as $R$-parity is conserved.
The neutralino LSP has been studied as a DM candidate 
in both the supersymmetric SM (MSSM) and its singlet 
extensions \cite{mssm,emssm,singlino}.

We may consider both the leptogenesis and the explanation of the DM abundance 
in supersymmetric models, simultaneously. 
In that case, the out-of-equilibrium decay of thermal heavy neutrinos 
can generate the sufficient baryon number asymmetry only if the reheating
temperature $T_R$ is high enough such as $10^8$~GeV or more \cite{leptg2}. 
If such a high reheating temperature is required, however, we confront the 
serious gravitino problem \cite{gravitino,newgravi}.
Various trials to overcome this difficulty have been done by searching
scenarios to enhance the $CP$ asymmetry or lower the required reheating
temperature \cite{maxcp,resonant,softlept}.

Recently, radiative neutrino mass models gather attentions as
candidates for physics at a TeV scale \cite{z2,nonsusy2,radmodel}.
The right-handed neutrinos and other new fields 
are assumed to have the masses of $O(1)$ TeV 
in many of such models. 
In this framework the small neutrino masses can be generated 
since the Dirac neutrino masses are assumed to be forbidden at tree-level 
by some symmetry such as $Z_2$ \cite{z2}.
Since the lightest one with this odd parity is stable because of this 
symmetry, it can be one of the DM candidates \cite{nonsusy2}.
This situation is the same as the LSP in the $R$-parity conserved MSSM. 
Unfortunately, the leptogenesis based on the decay of a thermal 
right-handed neutrino with the $Z_2 $ odd parity to the DM can not generate 
the sufficient lepton number asymmetry since it is too light. 
To remedy this fault, a hybrid model using both the radiative 
seesaw and the ordinary seesaw is proposed in a nonsupersymmetric 
framework \cite{hybrid}.  However, if we try to make this hybrid model 
supersymmetric, we confront again the gravitino problem as long as 
we insist on the decay of the thermal right-handed neutrinos.
Thus, it seems to be difficult to reconcile
this type of supersymmetric radiative neutrino mass models 
with the thermal leptogenesis.   
  
In this paper we propose an alternative possibility 
for the leptogenesis in the supersymmetric radiative neutrino mass model,
which can closely relate both the origin of the neutrino masses and the DM
abundance. We apply Affleck-Dine (AD) mechanism \cite{ad} 
based on a famous flat direction $LH_u$ to the model.
Since the neutrino mass generation is irrelevant to this flat direction
in this model, the model shows different features from the
ones found in the previous articles \cite{adap,others}.
We estimate the baryon number asymmetry generated through this
leptogenesis based on the AD mechanism and also the constraint on the
reheating temperature.

The remaining parts are organized as follows.
In section 2 we address the model and the flat direction considered here. 
In section 3 we estimate the baryon number asymmetry generated through
the evolution of this flat direction.
We show that the model can give us a simple and consistent picture 
for the explanations of the neutrino masses, the DM abundance, and 
the baryon number asymmetry. 
In section 4 we summarize the paper.

\section{A flat direction in the model}
\subsection{The radiative neutrino mass model}
We consider an extension of the MSSM with three singlet chiral 
superfields $N_i$, two extra 
doublet chiral superfields $\eta_u$ and $\eta_d$,
and also an additional singlet chiral superfield 
$\phi$ \cite{fks,sty}.\footnote{A similar supersymmetric model considered in
a different context can be found in \cite{e6}.} 
Lepton number $L$ is assigned to these superfields as
$L(N_i)=L(\phi)=0$ and $L(\eta_u)=-L(\eta_d)=-1$.
All their scalar components are assumed to have no vacuum 
expectation values.
We introduce a $Z_2$ symmetry in addition to the ordinary $R$-parity.
The charge assignment for these discrete symmetries are summarized in Table~1.
As a result of these discrete symmetries, we have two DM components, 
that is, the lightest ordinary neutralino $\chi$ and the lightest 
$Z_2$ odd neutral field.
In the following discussion, we assume that the lightest $N_i$ is
lighter than $\eta_{u,d}$ and $\phi$ among the $Z_2$ odd fields. 
We require that the gauge invariant superpotential constructed 
by these chiral superfields 
should also be invariant under the $R$-parity and also the $Z_2$ symmetry.

\begin{table}[t]
\begin{center}
\begin{tabular}{|c|ccccccccccc|} \hline
$\Psi$ &$Q_i$ & $\bar U_i$ & $\bar D_i$ & $L_i$ & $\bar E_i$ &$H_u$&$H_d$ 
& $N_i $ & $\eta_u $ & $\eta_d$ & $\phi $  \\ \hline\hline
$R$  & $-$ & $-$ &$-$& $-$ & $-$ & $+$&  $+$ &  $+$ &  $-$ 
& $-$& $-$ \\ \hline
$Z_2$ & $+$ & $+$ &$+$& $+$ & $+$ &  $+$&  $+$& $-$ &  $-$ 
& $-$& $-$  \\ \hline
$L$ & $0$ & $0$ &$0$& $+1$ & $-1$ &  $0$&  $0$& $0$ &  $-1$ 
& $+1$& $0$  \\ \hline
\end{tabular}

\caption{\footnotesize Matter contents and their quantum number. 
$R$ stands for the ordinary $R$ parity and $Z_2$ is a new parity which forbids 
the Dirac neutrino masses at  tree level.
$L$ is the lepton number.}
\end{center}
\end{table}

The invariant renormalizable superpotential is expressed as
\begin{eqnarray}
W&=&h_{ij}^U Q_{i} \bar U_{j}  H_u
+ h_{ij}^DQ_{i} \bar D_{j} H_d
+ h_{i}^EL_{i} \bar E_{i} H_d
+\mu_H H_u H_d, \nonumber \\
&+&h_{ij}^NL_{i} N_{j} \eta_u
+\lambda_u\eta_u H_d \phi
+\lambda_d\eta_d H_u \phi
+\mu_\eta\eta_u\eta_d
+ \frac{1}{2}M_i N_i^2 
+\frac{1}{2}\mu_\phi\phi^2,
\label{superpot}
\end{eqnarray} 
where all couplings and mass parameters are supposed to be real, for
simplicity. 
The MSSM superpotential is contained in the first line.
The second line includes the additional terms to the MSSM. 
They are relevant to the radiative neutrino mass generation.
Following the lepton number assignment to the fields shown in Table~1, 
the lepton number is violated only through the Yukawa couplings 
$\lambda_u\eta_uH_d\phi$ and $\lambda_d\eta_dH_u\phi$.
Since detailed discussion on the neutrino mass generation and 
the DM abundance in this model can be found in \cite{sty},
we do not repeat it. Here we review only some important features
of the model for the following discussion on the leptogenesis based
on the Affleck-Dine mechanism.

The neutrino masses are generated through the one-loop diagrams
 generated by the terms in the second line of eq.~(\ref{superpot}). 
An interesting point in this mass generation is that
the tri-bimaximal MNS matrix is automatically realized if the simple 
flavor structure is assumed for the neutrino Yukawa couplings such as
\begin{equation}
h_{ei}^N=0, \quad h_{\mu i}^N=h_{\tau i}^N ~(i=1,2), \qquad
h_{e3}^N=h_{\mu 3}^N=-h_{\tau 3}^N.
\label{yukawa}
\end{equation}
Moreover, if we consider that chiral superfield $\phi$ is much
heavier than the chiral superfields $N_i$ and $\eta_{u,d}$, the dominant
contribution to the neutrino masses takes a very simple form 
\begin{equation}
{\cal M}_\nu=\left(
\begin{array}{ccc}
0 & 0 & 0\\ 0 & 1 & 1 \\ 0 & 1 & 1 \\ \end{array}\right)
(h_{\tau 1}^2\Lambda_1+h_{\tau 2}^2\Lambda_2)
+\left(
\begin{array}{ccc}
1 & 1 & -1\\ 1 & 1 & -1 \\ -1 & -1 & 1 \\ \end{array}\right)
h_{\tau 3}^2\Lambda_3.
\label{nmass}
\end{equation}
The scales for the neutrino masses are determined by $\Lambda_i$
which is defined as
\begin{eqnarray}
&&\Lambda_i=\frac{\bar\lambda v^2 M_i}
{16\pi^2}\Big(g(M_i,m_{\eta +})-g(M_i,m_{\eta -})\Big), 
\qquad \bar\lambda=\frac{\lambda_u\lambda_d\tan\beta}{1+\tan^2\beta},
\nonumber \\
&&g(m_a,m_b)=\frac{m_b^2 -m_a^2+m_a^2\ln(m_a^2/m_b^2)}{(m_b^2-m_a^2)^2},
\label{mscale}
\end{eqnarray}
where $\langle H_u^0\rangle=v\sin\beta$ and $\langle H_d^0\rangle=v\cos\beta$.
$m_{\eta\pm}^2$ are the mass eigenvalues of the neutral scalar 
components of $\eta_{u,d}$, which are defined as 
$m_{\eta\pm}^2=\mu_\eta^2+m_0^2\pm B\mu_\eta$ by using supersymmetry
breaking parameters $m_0^2$ and $B$.
As long as $\mu_\eta$ and $M_i$ are assumed to be $O(1)$~TeV 
and $\bar\lambda$ is sufficiently suppressed as $\bar\lambda=O(10^{-8})$,
this neutrino mass matrix can explain the neutrino oscillation data
consistently with both constraints from the lepton flavor violating 
processes and the DM abundance \cite{sty}. 
In that case, since the Yukawa couplings $\lambda_u$ and $\lambda_d$ take very
small values of $O(10^{-4})$ for the $O(1)$ neutrino
Yukawa couplings $h^N_{ij}$, 
the lepton number violation in eq.~(\ref{superpot}) is found to be 
largely suppressed. 

Here it is useful to give a remark on the parameters in eq.~(\ref{superpot}).
In the above review of the neutrino masses, one might consider 
that a lot of ad hoc assumptions have made for the coupling constants 
and the mass parameters. 
However, as discussed in \cite{sty}, they could be justified if we suppose
to embed the $Z_2$ symmetry in an anomalous U(1) symmetry. In that case 
the superpotential $W$ is considered 
as an effective one induced from nonrenormalizable
interaction terms as a result of the spontaneous breaking of the
anomalous U(1) symmetry due to vacuum expectation values (VEVs) 
of singlet fields $\Sigma_+$ and $\Sigma_-$ at high energy regions. 
The coupling constants and the mass parameters in
the superpotential $W$ are expressed by using $\langle\Sigma_+\rangle$
or $\langle\Sigma_-\rangle$ as
\begin{eqnarray}
&&h_{ijk}=y_{ijk}\left(\frac{\langle\Sigma_\pm\rangle}{M_{\rm
		  pl}}\right)^{n_{ijk}}, \quad
n_{ijk}=-\frac{X_i+X_j+X_k}{X_{\Sigma_\pm}}\quad 
{\rm for}~~h_{ijk}\Psi_i\Psi_j\Psi_k, \nonumber \\
&&\mu_{ij}=y_{ij}M_{\rm pl}
\left(\frac{\langle\Sigma_\pm\rangle}{M_{\rm pl}}\right)^{n_{ij}}, \quad
n_{ij}=-\frac{X_i+X_j}{X_{\Sigma_\pm}} \quad 
{\rm for}~~\mu_{ij}\Psi_i\Psi_j,
\label{effp} 
\end{eqnarray}
where $X_i$ stands for the anomalous U(1) charge of the chiral
superfield $\Psi_i$. 
The coupling constants $y_{ijk}$ and $y_{ij}$ in the 
nonrenormalizable interaction terms in the original superpotential 
at high energy regions are supposed to be $O(1)$ naturally.
If the anomalous U(1) charge is assigned appropriately and 
the singlet scalars $\Sigma_\pm$ obtain favorable VEVs, 
these VEVs cause the hierarchical structure in the Yukawa couplings
of quarks and leptons, which realizes the favorable mass 
eigenvalues and mixing 
angles. Moreover, several parameters including $\lambda_{u,d}$ in the
superpotential $W$ are properly suppressed through eq.~(\ref{effp}). 
Such examples can be found in \cite{sty}.
We adopt this picture and use the example
given there. This example gives the following values for the parameters
relevant to the present discussion:
\begin{equation}
h^N_{ij}=O(1), \quad M_i,~\mu_\eta=O(1)~{\rm TeV}, \quad 
\lambda_{u,d}=O(10^{-4}), \quad \mu_\phi=O(10^8)~{\rm TeV}.  
\end{equation}
 
\subsection{A flat direction}
We now consider a flat direction of this model which is defined by a
single complex field $\varphi$ as
\begin{equation}
L_i=\left(\begin{array}{c} \varphi \\ 0 \\
\end{array}\right),  \qquad
H_u=\left(\begin{array}{c} 0 \\ \varphi \\
\end{array}\right),
\end{equation} 
where the scalar components of other chiral superfields are fixed to be zero.
The AD mechanism based on this flat direction and others has been studied 
in the MSSM and its extensions \cite{adap,others,nonpert}.
In such studies the flat direction is closely related to the 
neutrino masses. In particular, the lightest neutrino
mass is constrained by the relation to the reheating temperature.
This aspect can be changed in this model since the relevant operate
$L_iH_u$ has nothing to do with the neutrino mass generation 
as discussed above.

The flat direction is lifted by a nonrenormalizable interaction 
and also both supersymmetry breaking terms due to hidden sector
dynamics and finite energy density of an inflaton field.
As a result, the initial value of $\varphi$ is fixed and $\varphi$
evolves following the potential minimum determined by the 
evolution of the inflaton.   
As such a nonrenormalizable superpotential consistent with the imposed
symmetry discussed above, we find
\begin{equation}
W_{\rm nr}=\frac{\xi}{M}(L_iH_u)^2,
\end{equation}
where $M$ and $\xi$ can be determined by the symmetry imposed on the model.
In fact, if the model is considered to be invariant under the anomalous 
U(1) symmetry as discussed at the end of the previous part, $M=M_{\rm pl}$ and 
$|\xi|=O(10^{-6})$ may be expected. 
We use these in the following discussion.
It should be noted that the small value of $\xi$ is naturally 
realized in this picture. 

The scalar potential for $\varphi$ is induced by $W_{\rm nr}$ and also
by the above mentioned supersymmetry breaking effects.
The latter one is induced by the hidden sector dynamics 
which is characterized by the gravitino mass $m_{3/2}$ of $O(1)$~TeV  
and also the inflaton finite energy density which is characterized 
by the Hubble parameter $H$ \cite{adap}.
The feature of the AD mechanism is determined by this scalar potential
at the inflation era characterized by $H=H_I$ and also 
at the period after the inflation, that is, 
from the time when the inflation ends to the time $H\simeq m_{3/2}$ 
when $\varphi$ is expected to start moving toward a true minimum 
of the potential. 
This scalar potential is expressed as
\begin{equation}
V(\varphi,\varphi^\ast)=(m_\varphi^2-cH^2)|\varphi|^2
+\left(\frac{Am_{3/2}+aH}{M_{\rm pl}}
\xi\varphi^4 +{\rm h.c.}\right)+\frac{|\xi|^2}{M_{\rm pl}^2}|\varphi|^6,
\label{scalarp}
\end{equation}
where $m_\varphi^2=|\mu_H|^2+m_0^2$ which is considered to be $O(1)$~TeV.  
All the dimensionless constants $A$, $a$ and $c$ except for $\xi$ 
are considered to have values of $O(1)$.
We define the phases of $A$ and $a$  as 
$A\xi\equiv \tilde Ae^{i\theta_{A}}$
and $a\xi\equiv \tilde ae^{i\theta_{a}}$.
In this scalar potential, the lepton number is violated by $\Delta L=2$ 
through the terms in the parenthesis. 
Moreover, the same terms also violate the $CP$ invariance through the phases
$\theta_{A}$ and $\theta_{a}$.

If both $H>m_\varphi$ and $c>0$ are satisfied\footnote{If 
K\"ahler potential satisfies a suitable condition, the supergravity 
scalar potential can induce this kind of
supersymmetry breaking term corresponding to the negative squared mass. 
Such an example in case of the hidden 
sector supersymmetry breaking can be found in \cite{sz}, for example.}, 
the scalar potential (\ref{scalarp}) has a nontrivial minimum at
$\varphi_0(H)=|\varphi_0(H)|e^{i\theta_{\varphi_0}(H)}$ 
where $|\varphi_0(H)|$ is determined as a function of $H$ as 
\begin{equation}
|\varphi_0(H)|=\frac{M_{\rm pl}}{\sqrt 3|\xi|}
\left[\frac{2\tilde aH}{M_{\rm pl}}+
\left\{\left(\frac{2\tilde aH}{M_{\rm pl}}\right)^2
+\frac{3c|\xi|^2H^2}{M_{\rm pl}^2}\right\}^{1/2}\right]^{1/2}
\simeq \left(\frac{M_{\rm pl}H}{|\xi|}\right)^{1/2}, 
\label{pmin1}
\end{equation}
and the initial value of $\theta_{\varphi_0}$ can be expressed as
\begin{equation}
\theta_{\varphi_0}(H_I)=\frac{(2n+1)\pi}{4}-\frac{\theta_{a}}{4}, 
\label{pmin2}
\end{equation}
where $n$ is an integer.
The energy density of the universe is considered to be dominated 
by the inflaton during the evolution of $\varphi$. 
Thus, following the inflaton motion, the value of the Hubble parameter 
$H$ changes to induce the shift of the potential minimum $\varphi_0(H)$. 
The field $\varphi$ follows this minimum. Once the universe reaches the
time at which $H<m_\varphi$ is satisfied, a true minimum of the
potential $V(\varphi)$ appears at $\varphi=0$ and $\varphi$ starts the
motion toward this minimum with or without rotating around $\varphi=0$.    

The $\theta_\varphi$ dependence of the potential $V(\varphi)$ 
changes from $\cos(\theta_{a}+4\theta_\varphi)$ to
$\cos(\theta_{A}+4\theta_\varphi)$ when the universe changes over 
from the period $H>m_{3/2}$ to the period $H<m_{3/2}$.
Thus, during this transient time
torque is generated for the motion of $\varphi$ 
as long as $\theta_{A}$ is not equal to $\theta_{a}$.
If $|\theta_A-\theta_a|$ takes a larger value, the larger torque can be
caused and $\varphi$ could start rotating around $\varphi=0$.  
Since the lepton number density $n_L$ stored in the $\varphi$ configuration 
is expressed as $n_L=-2\dot{\theta}_\varphi|\varphi|^2$, 
the substantial lepton number is expected to be generated through this 
evolution of $\theta_\varphi$. 
In the next section, we estimate this induced lepton number by studying the 
evolution of the flat direction $\varphi$.   

\section{Leptogenesis based on the AD mechanism}
We follow the negative squared mass scenario given in \cite{adap} and then 
$c>0$ is assumed as in the above discussion here. 
In this case, the potential minimum during the 
inflation exists at $\varphi_0$ given in eqs.~(\ref{pmin1}) and (\ref{pmin2}).
Here we use $H_I\simeq 10^{14}$~GeV as the Hubble parameter during the
inflation, which is required by the density 
perturbation found in the CMB anisotropy observation.  
After this inflation period, the evolution of $\varphi$ is described
by the equation of motion which can be expressed as
\begin{equation}
\frac{d^2\varphi}{dt^2}
+3H\frac{d\varphi}{dt}+\frac{d V(\varphi,\varphi^\ast)}{d\varphi^\ast}=0,
\label{eqmphi}
\end{equation}
where the potential $V(\varphi,\varphi^\ast)$ is given in eq.~(\ref{scalarp}).
We are interested in the evolution of $\varphi$ during the period 
relevant to the lepton number generation. 
This period is characterized by
$H\sim m_{3/2}$. We suppose that the Hubble parameter $H$ is 
larger than the inflaton decay width $\Gamma_I$ in this period here. 
This is the case
if we confine our study to the case with the reheating temperature 
$T_R<10^{10}$~GeV. In this case
the universe is dominated by the matter due to the coherent oscillation
of the inflaton. Thus, we can use $H=\frac{2}{3t}$ in eq.~(\ref{eqmphi}).

Now we introduce a dimensionless Hubble parameter $x=\frac{H}{H_I}$ to
rewrite eq.~(\ref{eqmphi}) as its differential equations for the
dimensionless fields $\phi_{R,I}(x)$ which are defined as  
$\varphi(x)\equiv \frac{|\varphi_0(H_I)|}{\sqrt 2}(\phi_R(x)+i\phi_I(x))$. 
Thus, eq.~(\ref{eqmphi}) can be expressed as
\begin{equation}
\frac{d^2\phi_R}{dx^2}
+\frac{4}{9x^4}\frac{\partial V}{\partial\phi_R}=0, \qquad
\frac{d^2\phi_I}{dx^2}
+\frac{4}{9x^4}\frac{\partial V}{\partial\phi_I}=0,
\label{eqmvarphi1}
\end{equation}
where $\partial V/\partial\phi_R$ and $\partial V/\partial\phi_I$ are
given by
\begin{eqnarray}
&&\frac{\partial V}{\partial\phi_R}=
\left(\frac{m_{3/2}^2}{H_I^2}-cx^2\right)\phi_R 
+\frac{2m_{3/2}|\Phi_0|^2}{M_{\rm pl}H_I^2}
\Big(\tilde A\cos\theta_{A}+\frac{\tilde aH_Ix}{m_{3/2}}
\cos\theta_{a}\Big)
\Big(\phi_R^2-3\phi_I^2\Big)\phi_R \nonumber \\
&&-\frac{2m_{3/2}|\Phi_0|^2}{M_{\rm pl}H_I^2}
\Big(\tilde A\sin\theta_{A}+\frac{\tilde aH_Ix}{m_{3/2}}
\sin\theta_{a}\Big)
\Big(3\phi_R^2-\phi_I^2\Big)\phi_I
+\frac{3|\xi|^2|\Phi_0|^4}{4M_{\rm pl}^2H_I^2}
\Big(\phi_R^2+\phi_I^2\Big)^2\phi_R, \nonumber 
\end{eqnarray}
\begin{eqnarray}
&&\frac{\partial V}{\partial\phi_I}=
\left(\frac{m_{3/2}^2}{H_I^2}-cx^2\right)\phi_I
-\frac{2m_{3/2}|\Phi_0|^2}{M_{\rm pl}H_I^2}
\Big(\tilde A\cos\theta_{A}+\frac{\tilde aH_Ix}{m_{3/2}}
\cos\theta_{a}\Big)
\Big(3\phi_R^2-\phi_I^2\Big)\phi_I \nonumber \\
&&-\frac{2m_{3/2}|\Phi_0|^2}{M_{\rm pl}H_I^2}
\Big(\tilde A\sin\theta_{A}+\frac{\tilde aH_Ix}{m_{3/2}}
\sin\theta_{a}\Big)
\Big(\phi_R^2-3\phi_I^2\Big)\phi_R
+\frac{3|\xi|^2|\Phi_0|^4}{4M_{\rm pl}^2H_I^2}
\Big(\phi_R^2+\phi_I^2\Big)^2\phi_I, \nonumber \\
\end{eqnarray}
where we use the definition $\Phi_0=|\varphi_0(H_I)|$.

The behavior of $\phi_{R,I}$ is found by solving these equations 
numerically. 
Unfixed free parameters in these equations are $\theta_{A}$ and
$\theta_{a}$ only. The initial value of $\theta_{\varphi}$ at $x=1$ is
determined by $\theta_{a}$ as found from eq.~(\ref{pmin2}).
Although only $\theta_{A}$ is relevant to the potential
minimum at $x~{^<_\sim}~\frac{m_{3/2}}{H_I}(\equiv x_i)$, 
$\theta_{\varphi}$ could have any values there
since the global potential minimum should be realized at $\varphi=0$.
On the other hand, since $\tilde A=\tilde a=|\xi|$ is supposed here, 
both $\theta_{A}$ and $\theta_{a}$ play an equal role in the potential at the
transient period $x\sim x_i$. 
This could make a local potential minimum appear at $\varphi\not=0$ 
depending on the value of $\theta_{A}$.
Since such a local minimum appears at a separated place for a larger 
$|\theta_A-\theta_a|$, the larger torque could be induced to make 
$\dot\theta_\varphi$ larger as mentioned before. 
In that case, sufficient lepton number could be generated in the flat
direction and $\varphi$ is expected to rotate around $\varphi=0$ at
$x\ll x_i$.
On the other hand, if sufficient torque for the motion of $\varphi$ 
is not induced because of a small $|\theta_A-\theta_a|$ value,
the lepton number might not be sufficiently generated in the $\varphi$
and also the $\varphi$ might not show the rotating motion 
around $\varphi=0$ at $x<x_i$.   
In such a case, the generated lepton number might not be released into
the plasma at the appropriate time for leptogenesis. 
This could occur since the large mass of fields induced by $\varphi_0$ 
prohibits the evaporation of the flat direction into such fields.
In order to generate the lepton number in the plasma, $\varphi$ has to
store the sufficient lepton number when $\varphi$ starts
the oscillation around $\varphi=0$.
We need to examine these points through the numerical study.  

\input epsf
\begin{figure}[t]
\begin{center}
\epsfxsize=6cm
\leavevmode
\epsfbox{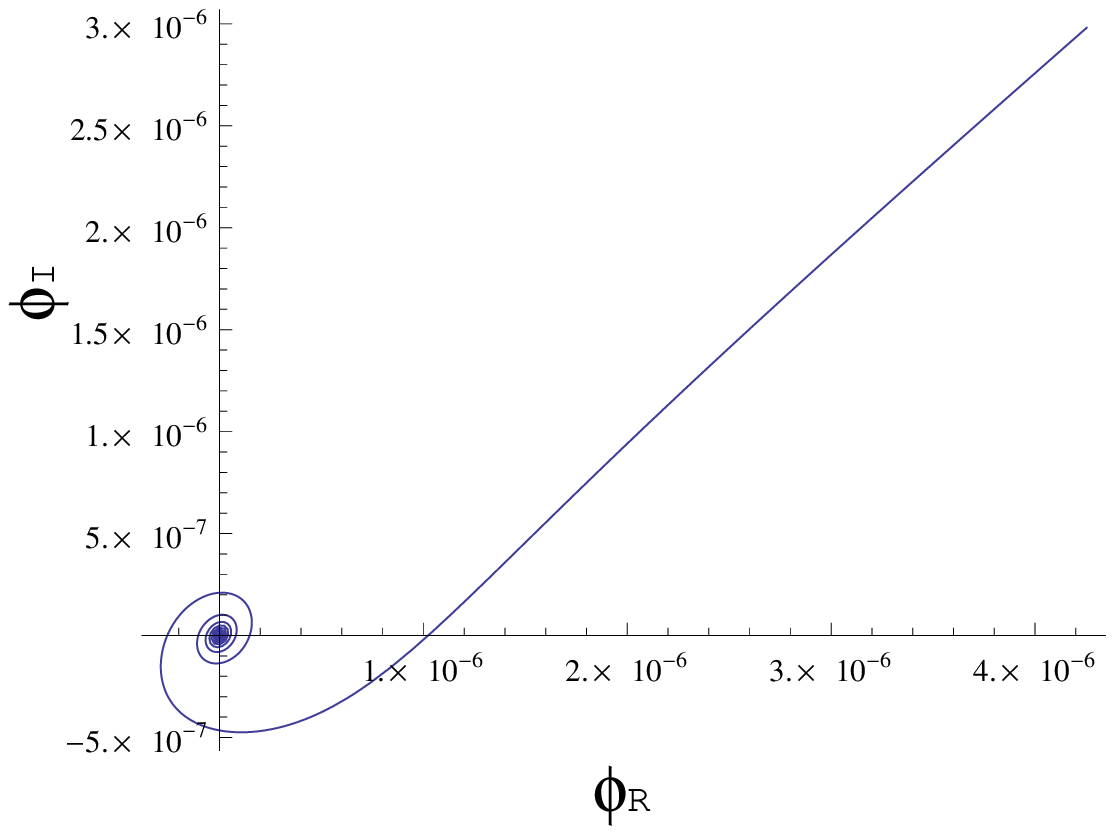}
\hspace*{6mm}
\epsfxsize=7cm
\leavevmode
\epsfbox{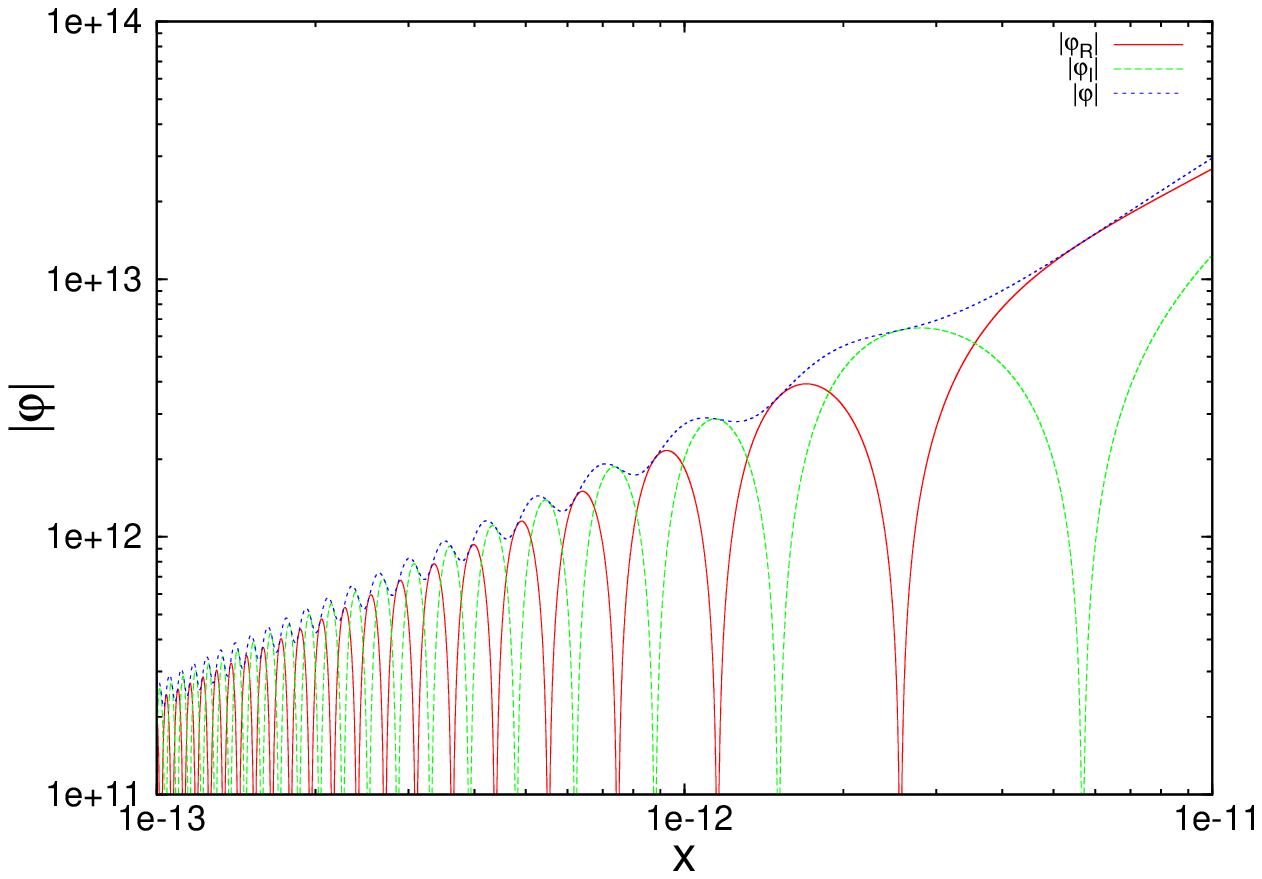}
\vspace*{2mm}
\epsfxsize=6cm
\leavevmode
\epsfbox{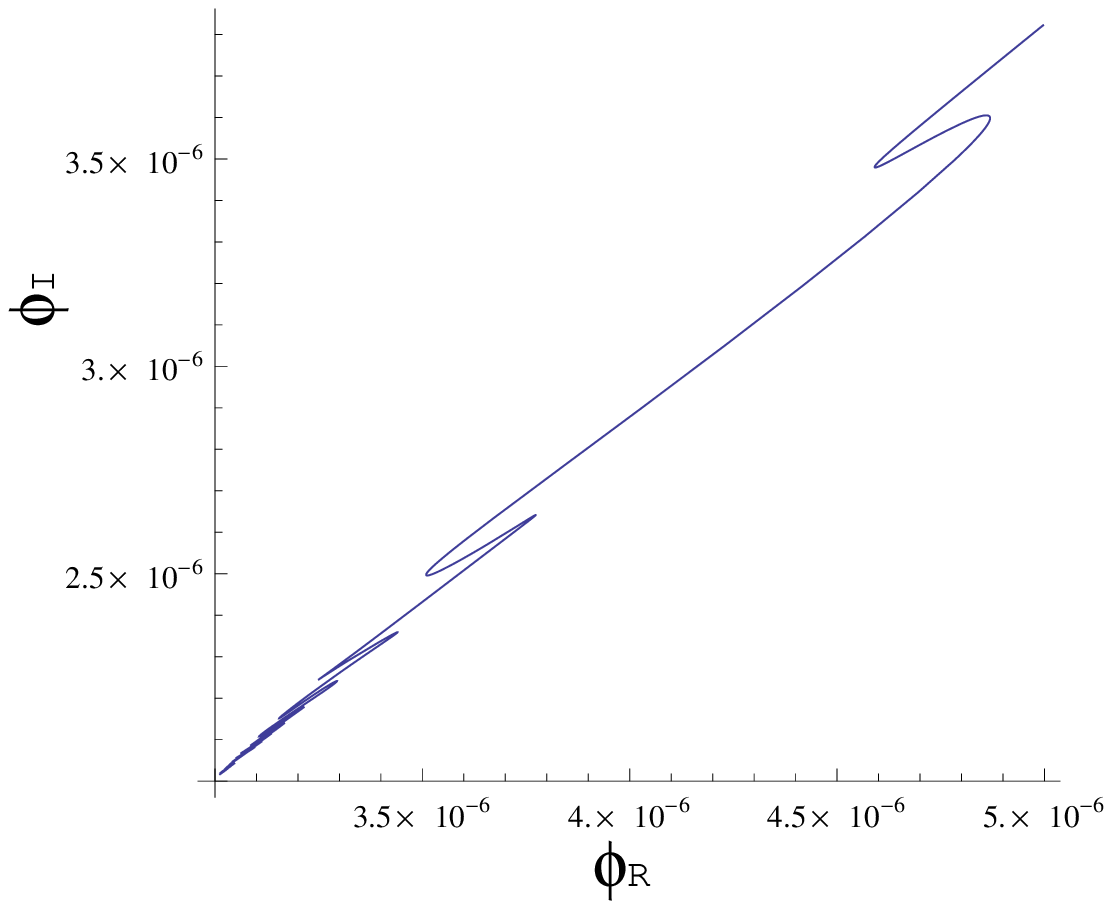}
\hspace*{6mm}
\epsfxsize=7cm
\leavevmode
\epsfbox{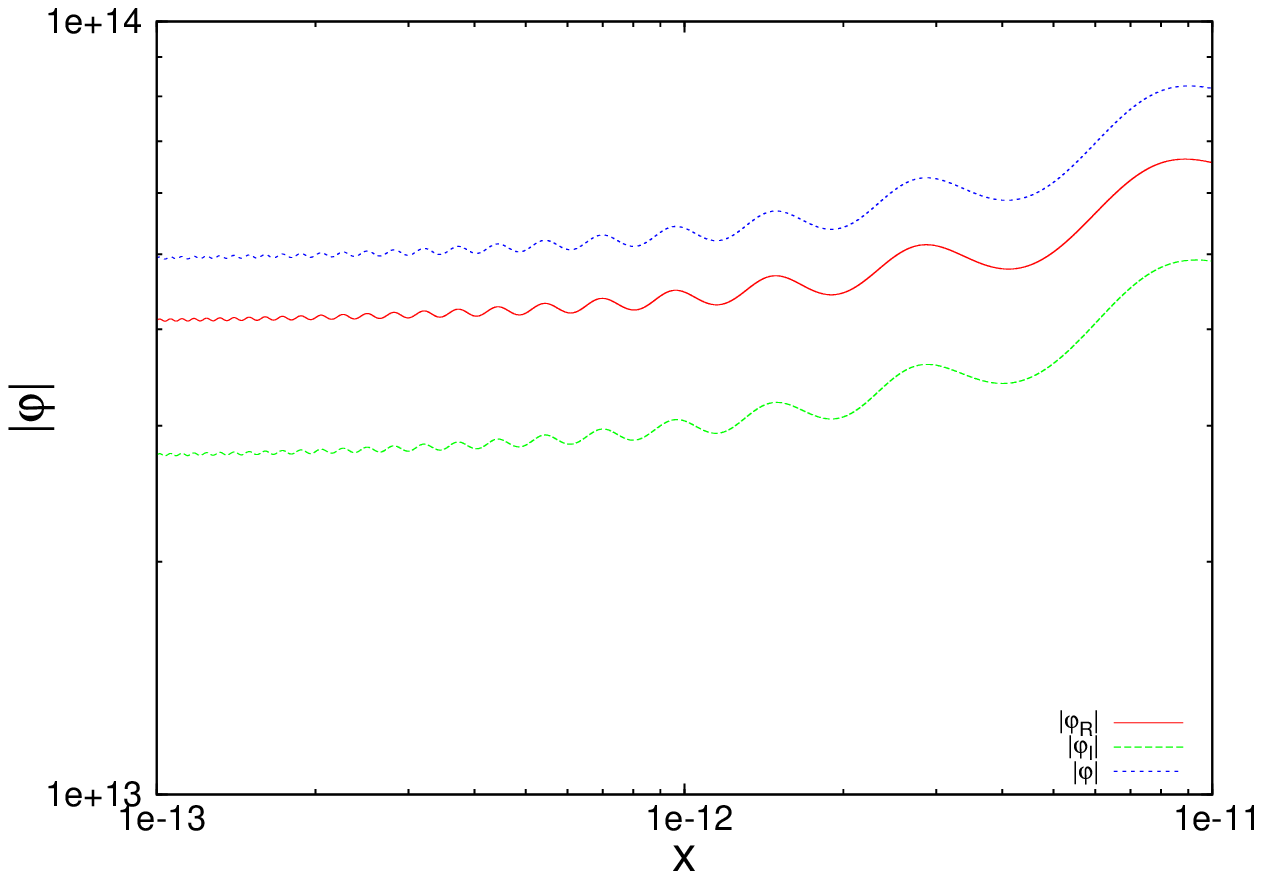}
\end{center}

\vspace*{-3mm}
{\footnotesize {\bf Fig.~1}~~Time evolution of the flat direction
 $\varphi$ for $\theta_A=\pi$ (above) and $\frac{\pi}{4}$
 (below). The trajectory in the $(\phi_R, \phi_I)$ plane is
 plotted in the left figure. In the right figure the evolution 
of the flat direction $|\varphi|$ (a blue dotted line)
 and its real and imaginary part $|\frac{\Phi_0}{\sqrt 2}\phi_R|$ (a red
 solid line), $|\frac{\Phi_0}{\sqrt 2}\phi_I|$ (a green dotted line) 
are plotted as functions of the dimensionless Hubble parameter $x$ in a
 region $10^{-2}x_i\le x\le x_i$.}
\end{figure}
 
In the numerical study we fix the free parameters as 
$\theta_{A}=\pi,~\frac{\pi}{4}$ and $\theta_{a}=\frac{\pi}{8}$ 
as typical examples for a while.   
The evolution of $\varphi$ in these cases is shown in Fig.~1. 
In the left figures the trajectory of $\varphi$ 
is plotted in the $(\phi_R, \phi_I)$ plane.
We also plot $|\varphi|$, $\frac{\Phi_0}{\sqrt 2}|\phi_R|$ and 
$\frac{\Phi_0}{\sqrt 2}|\phi_I|$ as functions of the dimensionless 
Hubble parameter $x(\equiv\frac{H}{H_I})$ in the right figures.
In case of $\theta_A=\pi$, $|\varphi|$ starts the oscillation 
around the origin $\varphi=0$ and $\dot{\theta}_\varphi$ can have large 
values at $x~{^<_\sim}~x_i$ as expected.
Thus, the sufficient lepton number can be generated in the $\varphi$ condensate.
On the other hand, in case of $\theta_A=\frac{\pi}{4}$, $|\varphi|$
does not oscillate around $\varphi=0$.
As a result, few lepton number is generated and it tends to
decrease at the $x<x_i$ region.
These suggest that a large amount of lepton number is not 
be expected to be generated for the case with small values 
of $|\theta_A-\theta_a|$.

\begin{figure}[t]
\begin{center}
\epsfxsize=8cm
\leavevmode
\epsfbox{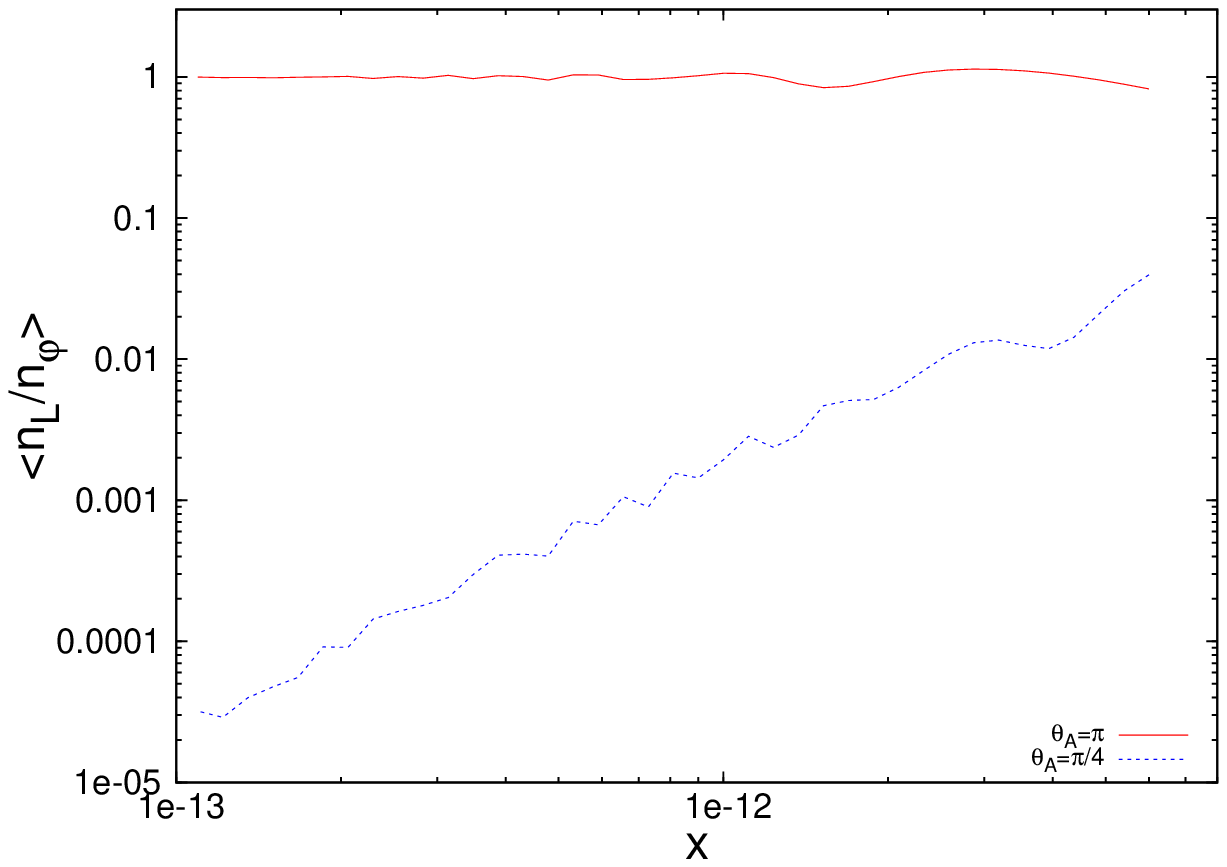}
\end{center}

\vspace*{-3mm}
{\footnotesize {\bf Fig.~2}~~The averaged lepton number
 $\langle\frac{n_L}{n_\varphi}\rangle_x$ generated in the plat
 direction $\varphi$ as a function of $x$.
Each line corresponds to the case $\theta_{A}=\pi$ and 
$\frac{\pi}{4}$, respectively. $\theta_a$ is fixed to $\frac{\pi}{8}$.}
\end{figure}

The averaged value of the generated lepton number $n_L$ at $x$ may be 
estimated by using these solutions as
\begin{equation}
\left\langle\frac{n_L}{n_\varphi}\right\rangle_x=
\frac{1}{\Delta_i+\Delta_f}\int^{x+\Delta_f}_{x-\Delta_i}dx
\left(\frac{-3H_Ix^2}{2m_\varphi}\right)
\left(\frac{\phi_R'\phi_I-\phi_R\phi_I'}{\phi_R^2+\phi_I^2}\right),
\label{average}
\end{equation}
where $\phi_{R,I}^\prime=\frac{d\phi_{R,I}}{dx}$ and $\Delta_i+\Delta_f$ 
is fixed so as to include a few oscillation cycles if $|\varphi|$ oscillates.
In Fig.~2 this averaged value of $\frac{n_L}{n_\varphi}$ is plotted 
as a function of $x$ for two typical values of $\theta_A$.
The figure shows that $\langle\frac{n_L}{n_\varphi}\rangle_x$ 
takes an almost constant value at the region $x~{^<_\sim}~x_i$ for 
$\theta_A=\pi$. On the other hand, the generated lepton number suddenly
decreases in case of $\theta_A=\frac{\pi}{4}$.
The generated lepton number can take largely different values 
depending on the relative value of $\theta_A$ and $\theta_a$ 
as mentioned above.

In the left figure of Fig.~3,\footnote{We can check that 
this quantity has the same behavior for the pair of 
$\theta_{A}$ and $\theta_{a}$ which satisfies
$\theta_{A}-\theta_{a}=\pi$. 
If we use this feature and Fig.~3, we can know the value of 
$\langle\frac{n_L}{n_\varphi}\rangle_x$ for any set of 
$\theta_A$ and $\theta_a$.} 
we show how the averaged value 
$\langle\frac{n_L}{n_\varphi}\rangle_{x=0.1x_i}$ 
depends on $\theta_{a}$ which determines the initial value 
of $\theta_{\varphi_0}$. 
The figure shows that $\langle\frac{n_L}{n_\varphi}\rangle$ can
have the values of $O(1)$ as long as the pair of $\theta_A$ and 
$\theta_{a}$ takes values in the suitable regions. 
Such regions are not so narrow as found from this figure.    

If we write the energy densities of the AD field and the inflaton 
as $\rho_\varphi$ and $\rho_I$ respectively, they can be expressed as 
$\rho_I=sT_R$ and $\rho_\varphi=m_\varphi n_\varphi$
by using the entropy density $s$ and the reheating temperature $T_R$.
Taking account of these relations and the fact that the 
inflaton dominates the energy of the universe,
 we obtain the ratio of the lepton number density to 
the entropy density at $H\sim m_{3/2}$ as
\begin{equation}
\left.\frac{n_L}{s}\right|_{H\sim m_{3/2}}\simeq\frac{n_L}{n_\varphi}
\frac{T_R}{m_\varphi}\frac{\rho_\varphi}{\rho_I}
\simeq \frac{n_L}{n_\varphi}\frac{T_R\varphi^2(x_i)}{m_{3/2}M_{\rm pl}^2},
\end{equation}
where we use $\rho_I\simeq (m_{3/2}M_{\rm pl})^2$ and 
$\rho_\varphi\simeq m_{3/2}^2\varphi^2$ at $H\simeq m_{3/2}$. 
Since the sphaleron transition can be in the thermal equilibrium after
the reheating ($H\le\Gamma_I$), 
it causes the reprocessing from the $B-L$ asymmetry to the $B$ 
asymmetry through the relation $B=\frac{1}{4}(B-L)$ 
in the present model.\footnote{In this derivation we suppose that the
lepton number violating interactions $\lambda_u\eta_uH_d\phi$ and 
$\lambda_d\eta_dH_u\phi$ are out-of-equilibrium. This point is discussed
below.} 
Thus, the generated baryon number asymmetry can be estimated 
as\footnote{The sign is not crucial here since we can find that the other 
pair of $\theta_{A}$ and $\theta_{a}$ can generate the same values
of $|\varphi(x)|$ and also 
$\left|\left\langle\frac{n_L}{n_\varphi}\right\rangle_x\right|$ 
with opposite sign from Fig.~3.} 
\begin{equation}
\left.\frac{n_B}{s}\right|_{x}=
\frac{1}{4}\left\langle\frac{n_L}{n_\varphi}\right\rangle_{x}
\frac{T_R\varphi^2(x)}{m_{3/2}M_{\rm pl}^2}
\simeq 4.3\times 10^{-10}\left(
\frac{\langle n_L/n_\varphi\rangle_{x}}{1}\right)
\left(\frac{\varphi(x)}{10^{13}~{\rm GeV}}\right)^2
\left(\frac{T_R}{10^5~{\rm GeV}}\right).
\label{basym}
\end{equation}
This formula and both figures in Fig.~3 suggest that the favorable
$\frac{n_B}{s}$ can be generated for suitable values of $\theta_a$,
which include the previously discussed 
example $\theta_A=\pi$ and $\theta_a=\frac{\pi}{8}$. 
We find that rather low reheating temperature $T_R\simeq 10^{5-6}$~GeV is
acceptable as long as $\varphi$ is evaporate to the plasma during 
$0.1x_i~{^<_\sim}~x~{^<_\sim}~x_i$. 
Such a reheating temperature $T_R$ is high enough for the 
sphaleron process to be in the thermal equilibrium. 
It is also sufficiently low to escape the gravitino problem.
Although $\langle\frac{n_L}{n_\varphi}\rangle$ seems to have much smaller
values than $O(1)$ at $0.1x_i~{^<_\sim}~x~{^<_\sim}~x_i$ for
$|\theta_a|~{^>_\sim}~1$ which includes the equivalent case to
the previous example $\theta_A=\frac{\pi}{4}$ and $\theta_a=\frac{\pi}{8}$,
$|\varphi(x)|$ takes larger values than $10^{13}$~GeV and then it
might seem to make $\frac{n_B}{s}$ be of $O(10^{-10})$ 
even for $T_R\simeq 10^{4-6}$~GeV. However, in that case, the generated
lepton number could not be successfully released into the plasma as
discussed below. 

\begin{figure}[t]
\begin{center}
\epsfxsize=7cm
\leavevmode
\epsfbox{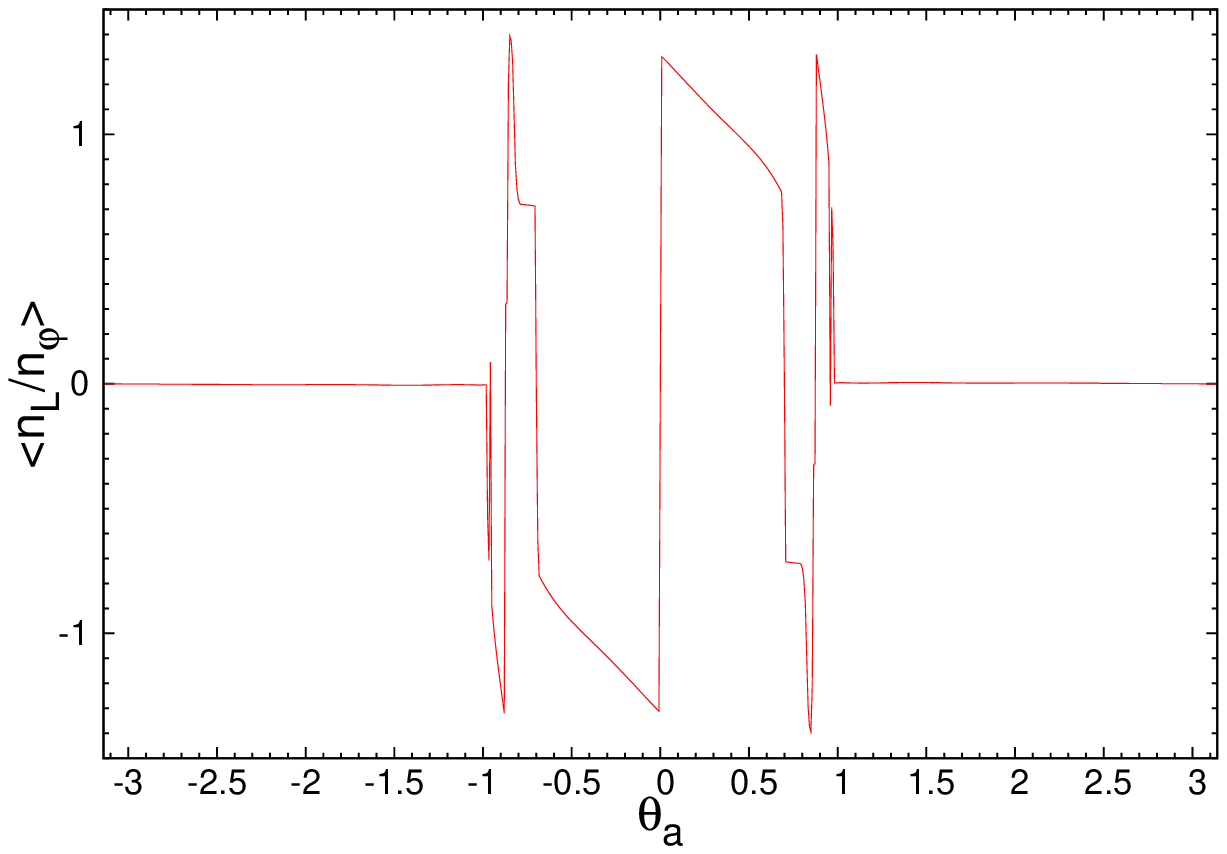}
\vspace*{5mm}
\epsfxsize=7cm
\leavevmode
\epsfbox{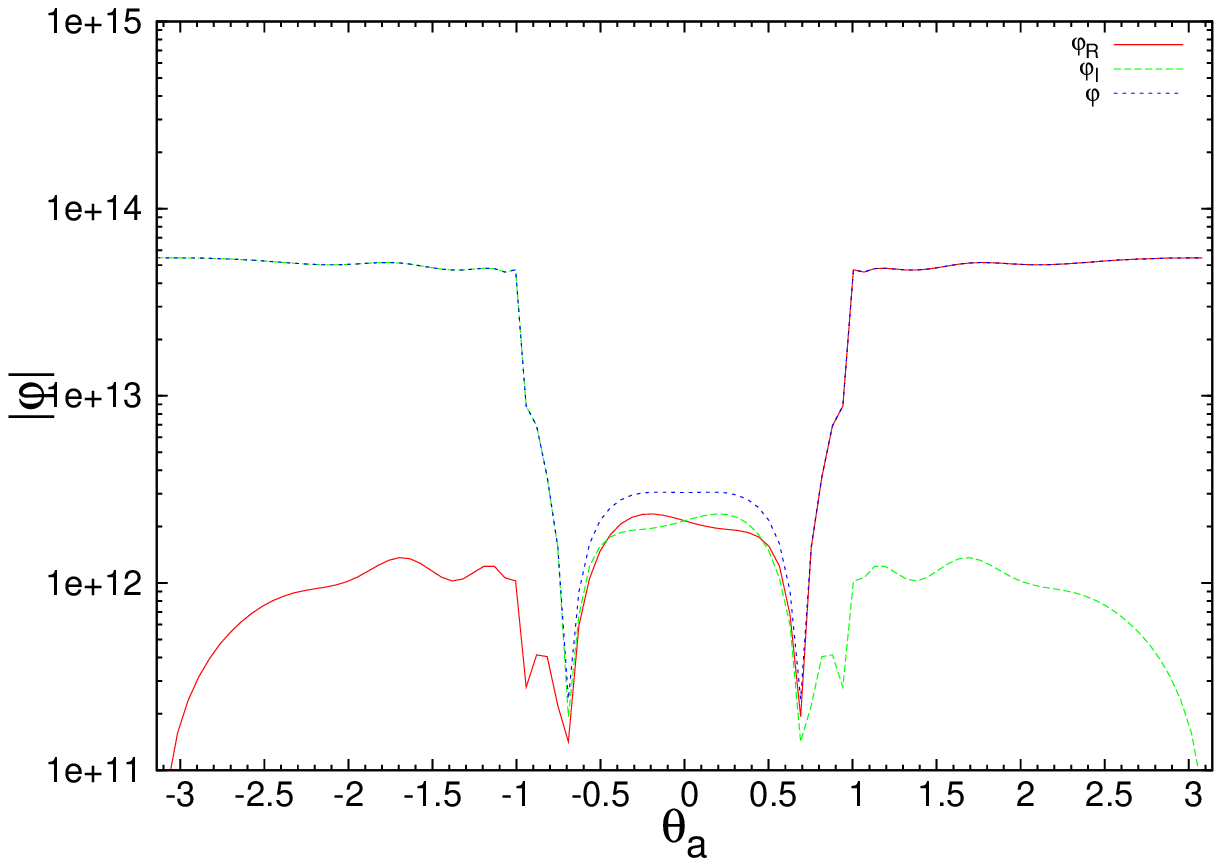}
\end{center}

\vspace*{-3mm}
{\footnotesize {\bf Fig.~3}~~The $\theta_{a}$ dependence 
of the generated lepton number $\langle\frac{n_L}{n_\varphi}\rangle_x$ 
and the value of $|\varphi(x)|$ at $x=0.1x_i$. In the right figure each line
 represents the same one as in the right figure of Fig.~1.
$\theta_{A}$ is fixed to $\pi$ here.}
\end{figure}

It is useful to present a remark on this reheating temperature $T_R$ here.
Since $|\xi|$ is strongly suppressed in this scenario,
the large value of $\varphi(x_i)$ can be realised.
It makes the value of $T_R$ required for the generation of the appropriate
baryon number asymmetry lower in comparison with the ordinary scenario 
based on the $LH_u$ direction \cite{adap}. 
It should also be noted that there is no constraint on the neutrino masses 
since the $LH_u$ direction is irrelevant to the neutrino masses
in this model.

Next, we need to examine the conditions which are required 
to identify the value of $\frac{n_B}{s}$ given by eq.~(\ref{basym}) 
with the observed baryon number asymmetry in the universe.
Eq.~(\ref{basym}) is estimated under the assumption that
all the lepton number generated at the time $H\sim m_{3/2}$ is
transformed to the baryon number asymmetry. This estimation gives the
correct answer only if the following conditions are satisfied:\\
(i)~ The evaporation of the flat direction 
due to both the decay and the scattering with the plasma yielded 
from the inflaton decay should be forbidden before the time 
$H\sim m_{3/2}$. Otherwise, the sufficient $\frac{n_L}{n_\varphi}$ can not be
generated. \\
(ii)~The lepton number violating interactions with the flat direction $\varphi$
should decouple during the period from $H\sim m_{3/2}$ to the sphaleron
decoupling time.

We examine the condition (i) at first.
If the relevant processes occur, $\Gamma>H$ and also the kinematical 
condition $\sum_i E_i >\sum_f m_f$ should be satisfied, 
where $\Gamma$ is the reaction rate of the relevant processes. 
$E_i$ and $m_f$ represent the energy of the fields included in the 
initial state and the mass of fields included in the final state, 
respectively.
The energy of the thermal plasma before the reheating is estimated 
by the temperature $T_r\simeq k_r(M_{\rm pl}HT_R^2)^{1/4}$ 
where $k_r=\left(\frac{72}{5\pi^2 g_\ast}\right)^{1/8}$ and 
$T_R$ is the reheating temperature \cite{ktbook}.
Since the scattering rate of $\varphi$ with this plasma is roughly
estimated as $\Gamma\simeq \alpha_y^2T_r$ where
$\alpha_y=\frac{y^2}{4\pi}$ and $y$ is a relevant coupling constant, 
$\Gamma~{^>_\sim}~H$ is satisfied for 
\begin{equation}
H~{^<_\sim}~\left(k_r^4\alpha_y^8M_{\rm pl}T_R^2\right)^{1/3}\sim 
m_{3/2}\left(\frac{y}{0.25}\right)^{16/3}
\left(\frac{T_R}{10^5~{\rm GeV}}\right)^{2/3.},
\label{decay}
\end{equation}
where we use $g_\ast\simeq 100$ as the relativistic degrees of freedom.
This means that $\varphi$ could evaporate through the exchange of 
top quark which has $y\simeq 1$ before the time $H\sim m_{3/2}$ if
$T_R\sim 10^5$~GeV is assumed. 
However, we need to note that the masses of the fields in the final state 
are induced as $y|\varphi_0(H)|$ through the interaction with the flat 
direction, where $y$ is a relevant coupling constant with the flat
direction $\varphi$. 
The measure for this kinematical condition is given by \cite{adap}
\begin{equation}
\left.\frac{y|\varphi_0(H)|}{T_r}\right|_{H\sim m_{3/2}}
\simeq\Big(\frac{(m_{3/2}M_{\rm pl})^{1/2}}{|\xi|T_R}\Big)^{1/2}
\sim~\frac{y}{3\times 10^{-6}}\left(\frac{10^5~{\rm GeV}}{T_R}\right)^{1/2}. 
\end{equation}
Even if we assume $y\sim 10^{-5}$ which corresponds to the electron case
and gives the severest condition,
$y|\varphi_0|>T_r$ is satisfied even for $T_R\sim 10^5$~GeV. 
From this discussion, we find that $\frac{n_L}{n_\varphi}$ can reach 
a suitable value before the evaporation of the flat direction 
keeping the consistency with the reheating temperature 
which should satisfy $T_R\ge 10^2$~GeV.

We also note that the flat direction can evaporate to the thermal
plasma before the sphaleron decoupling at $T\sim 10^2$~GeV.
Since the $\varphi$ starts the oscillation around the potential minimum
$\varphi_0=0$ for suitable values of $\theta_A$ and $\theta_a$ 
once $H<m_{3/2}$ is fulfilled, 
any contribution to the masses of the final states 
is not induced through the interaction with $\varphi$.
Fig.~1 gives such a concrete example. 
As long as such a situation is realized, we know from the discussion 
on eq.~(\ref{decay}) that the lepton number asymmetry stored in $\varphi$ is
released into the plasma immediately through the lepton number
conserving scattering processes.\footnote{Nonperturbative effects might
play an important role in the decay of the flat direction \cite{nonpert}.} 
Such processes contain $\varphi\tau\rightarrow\tilde\tau\nu_\tau$ and
$\varphi b\rightarrow\nu_\tau\tilde b$ through 
a Higgsino $\tilde H_d$ exchange. 

The lepton number violating interactions are given by the
$\lambda_{u,d}$ terms in eq.~(\ref{superpot}). 
The decay of the flat direction 
through the interaction $\lambda_d\eta_dH_u\phi$ 
is kinematically forbidden since
the mass of the singlet field $\phi$ is large enough.  
The rate of the $\varphi$ scattering through the $\phi$ exchange 
is proportional to $|\lambda_{u,d}|^4$. 
Since $|\lambda_{u,d}|$ is considered to be $O(10^{-4})$ as discussed in
the previous part, $\Gamma<H$ is satisfied throughout the relevant period.
Thus, the dangerous lepton number violating processes are irrelevant to the
present scenario and the condition (ii) is also fulfilled.
All the lepton number asymmetry induced in the flat direction at 
$H\sim m_{3/2}$ is distributed in the plasma and then it is converted 
to the baryon number asymmetry through the sphaleron process 
as discussed above.
Favorable features found in the AD mechanism for the $LH_u$ flat
direction in the MSSM are kept even for the rather low reheating
temperature such as $T_R\sim 10^{5}$~GeV in the present framework.  

\section{Summary}
The supersymmetric radiative neutrino mass model is an interesting 
extension of the SM. 
It can give a consistent explanation for both the origin of the small neutrino
masses and the DM abundance by relating them closely.
In this model, however, it seems to be difficult to generate the
required baryon number asymmetry in the universe through the ordinary
thermal leptogenesis.
The right-handed neutrinos with the masses of $O(1)$~TeV are too 
light to generate the sufficient lepton number asymmetry through 
their out-of-thermal equilibrium decay. 

In this paper we have considered the leptogenesis based on the AD
mechanism as an alternative possibility. 
We have applied the AD mechanism to the famous flat direction $LH_u$.
Since this flat direction is irrelevant to the neutrino mass generation
in this model, any constraint on the neutrino masses does not appear
in the relation to the AD mechanism.
This is largely different from the previous work. 
Our analysis shows that the model can produce the sufficient baryon 
number asymmetry through the AD mechanism based on this flat direction 
for the rather low reheating temperature such as $10^5$~GeV.   
It is interesting that the crucial parameters for this mechanism 
can be related to other parameters in the model by introducing the 
anomalous U(1) symmetry. 
We would like to stress that this symmetry is embedded in the present
model so as to give the consistent explanation for 
the problems which remain as unsolved ones in the SM, that is, 
the neutrino masses,
the dark matter abundance and also the baryon number asymmetry in the
universe. 
\vspace*{5mm}

\section*{Acknowledgement}
This work is partially supported by a Grant-in-Aid for Scientific
Research (C) from Japan Society for Promotion of Science (No.21540262)
and also a Grant-in-Aid for Scientific Research on Priority Areas 
from The Ministry of Education, Culture, Sports, Science and Technology 
(No.22011003).

\end{document}